\documentclass[a4paper,11pt]{article}
\usepackage{pos}
\usepackage{hyperref}
\usepackage{ulem} 

\title{QCD corrections for Pseudoscalar Higgs decay to 3 partons at higher orders in dimensional regulator }

\author*[a]{Pulak Banerjee}
\author[b]{Chinmoy Dey}
\author[c]{ M.C.Kumar}
\author[d]{ V. Ravindran}

\affiliation[a]{Department of Physics, School of Advanced Sciences, Vellore Institute of Technology (VIT),
Chennai Campus, Chennai 600127, Tamil Nadu, India.\\}

\affiliation[b]{Theoretical Physics Division, Physical Research Laboratory, Navrangpura, Ahmedabad 380009, India}

\affiliation[c]{Department of Physics, Indian Institute of Technology Guwahati, Guwahati-781039, Assam, India\\}

\affiliation[d]{The Institute of Mathematical Sciences, A CI of Homi Bhabha National Institute, Taramani, Chennai 600113, India\\}

\emailAdd{pulak.banerjee@vit.ac.in}
\emailAdd{chinmoy@prl.res.in}
\emailAdd{mckumar@iitg.ac.in}
\emailAdd{ravindra@imsc.res.in }

\abstract{In this contribution, we present our recent study on the second-order corrections of pseudo-scalar($A$) Higgs decay to three partons, at higher orders in the dimensional regulator. We have studied the 
one and two-loop amplitudes for processes, $A\to ggg$ and $A\to q\bar{q}g$ in the effective theory framework. Our results, after suitable crossings of the external momenta, are important ingredients for predicting the differential distribution of the pseudo-scalar Higgs boson in association with a jet at hadron colliders.}

\FullConference{17th International Symposium on Radiative Corrections: Applications of Quantum Field Theory to Phenomenology (RADCOR2025)\\
5-10 October 2025\\
Puri, India\\}

\begin{document}
\maketitle
\section{Introduction}
Ever since the discovery of the Higgs boson in 2012 at the Large Hadron Collider (LHC) \cite{ATLAS:2012yve,CMS:2012qbp}, establishing its CP properties and the measurement of its couplings to the Standard Model (SM) particles 
has gained significant importance. The gluon-fusion channel is the dominant mechanism for Higgs production at the LHC and receives large perturbative corrections. Consequently, to achieve the theoretical prediction in this channel, higher order corrections up to third order in pQCD have been computed \cite{Ravindran:2002dc,Harlander:2002vv,Ravindran:2003um,Anastasiou:2015vya,Anastasiou:2016cez}.
In spite of the success of the SM, it has few shortcomings in explaining some aspects like the origin of neutrino mass, Higgs hierarchy problem and unification of gauge couplings. Many models beyond the SM (BSM) have been proposed to address such issues, Supersymmetry 
being such a BSM scenario. In the context of Minimal Supersymmetric Standard Model (MSSM), after the electroweak symmetry breaking, the spectrum of scalar sector includes three neutral scalars ($h, H, A$) and two charged Higgs ($H^\pm$) scalars \cite{Fayet:1974pd}.
Out of the neutral scalars, the CP-odd pseudoscalar ($A$) has gained a lot of attention particularly in identifying whether the discovered Higgs boson is a CP-even or CP-odd state or if it has a mixture of these two components. Consequently, the search of CP-odd scalar in the collider experiments has become very crucial. In the perturbation theory, the production of the pseudo-scalar takes place via gluon fusion channel, similar to the case of the SM Higgs boson, and hence the associated higher order QCD corrections are also expected to be as large as those for the SM Higgs boson, 
The main difference being that the pseudo-scalar in the EFT has two different operators $O_G$ and $O_J$. 
The higher order corrections to such observables are non-trivial. The NLO corrections for pseudo-scalar Higgs in association with a jet can be found in \cite{deFlorian:1999zd,Ravindran:2002dc, Field:2002pb, Bernreuther:2010uw}, while the NNLO corrections for the same process have been recently presented in \cite{Kim:2024kaq}. 
The computation of higher order corrections beyond NNLO to these differential distributions becomes very non-trivial, however, attempts are going in this direction. One necessary ingredient at third order comes from the one loop virtual results expanded to $\epsilon^4$, the two loop results expanded to $\epsilon^2$, and three loop level results expanded to $\epsilon^0$ level. For the case of Higgs + 1 jet, the two-loop results expanded to $\epsilon^2$ have been recently computed in \cite{Gehrmann:2023etk}. In the context of pseudoscalar plus one jet, such results can be obtained by performing first the two-loop virtual computation for the pseudo-scalar decay to three partons up to $\epsilon^2$ and then using appropriate crossing symmetry relations will give the required result for the pseudo-scalar + 1 jet. Motivated by this, in this work, we perform the two-loop computation for the pseudo-scalar decay to three partons and present the results expanded to $\epsilon^2$ and present a brief numerical study of the finite pieces ($\epsilon^i$ ($i \geq 0$) for both $A \to ggg$ and $A \to q \bar{q} g$ cases at one-loop as well as two-loop level in QCD.
In the next sections, we briefly present the relevant theoretical framework, the calculation details along with the UV renormalization IR regularization, and finally we conclude.

\section{Theoretical details}
The coupling of a pseudo-scalar Higgs boson to heavy quarks occurs through the Yukawa interaction. Working within an effective theory, in the limit of infinite quark mass, the Lagrangian can be written as follows \cite{Chetyrkin:1998mw}:

\begin{align}
 {\cal L}^{A}_{\rm eff} = \Phi^{A}(x) \Big[ -\frac{1}{8}
  {C}_{G} O_{G}(x) - \frac{1}{2} {C}_{J} O_{J}(x)\Big]
  \label{eq:efflag}
\end{align}
 where $\Phi^{A}$ being the the pseudo-scalar Higgs boson. $C_G$ and $C_J$ are  Wilson 
 coefficients,
which multiply the two operators $O_{G}(x)$ and  $O_{J}(x)$. 
For details, see \cite{Banerjee:2024zho}. The Wilson coefficients $C_G$ and $C_J$ can be expanded in terms of the strong coupling constant, $a_s = \frac{g_s^2}{16 \pi^2}$.  The expression for $C_G$ and $C_J$  are as follows:
\begin{align}
  \label{eq:const}
  & C_{G} = a_{s} C_G^{(1)}
    \nonumber\\
  & C_{J} =  \left( a_s C_J^{(1)}  + a_s^2 C_J^{(2)}  +... \right) C_G \\
  & \text{with} ~C_G^{(1)} = -2^{\frac{5}{4}}G_F^{\frac{1}{2}}\text{cot}\beta, ~C_J^{(1)} = -C_F(\frac{3}{2}-3\ln\frac{\mu_R^2}{m_t^2}). \nonumber
\end{align}
The operators, $O_{G}(x)$ and  $O_{J}(x)$  are defined in terms of the Standard Model fermionic and gluonic fields. $O_{J}(x) $, being a chiral quantity, contains $\gamma_5$
and the Levi-Civita tensor $\varepsilon^{\mu\nu\rho\sigma}$. These quantities are not well defined for  $d \neq 4$ dimensions; it is essential to choose a proper definition of them. We shall follow the 
definition introduced by ’t Hooft and Veltman in the article \cite{tHooft:1972tcz}:
\begin{align}
\label{eq:g5}
  \gamma_5 = i \frac{1}{4!} \varepsilon_{\nu_1 \nu_2 \nu_3 \nu_4}
  \gamma^{\nu_1}  \gamma^{\nu_2} \gamma^{\nu_3} \gamma^{\nu_4} \,.
\end{align}

\section{Calculational framework}
For the decay channel process:
 \begin{equation}
  K(q) \rightarrow B(p_1) + C(p_2) + D(p_3)
 \end{equation}
where $\{B,C,D\} =  \{g,g,g\}$ or $\{B,C,D\} = \{q,\bar{q},g\}$. The Mandelstam invariants are defined 
as:
\begin{equation}
s = (p_1 + p_2)^2, t = (p_1 + p_3)^2, u = (p_2 + p_3)^2, \, q^2=Q^2.
\end{equation}
Here, $Q$ is the mass of pseudoscalar Higgs boson and they satisfy the conservation relation:
\begin{equation}
s+t+u= Q^2
\end{equation}
We shall work with dimensionless variables, $x=s/Q^2$, $y=u/Q^2$ and $z=t/Q^2$; the kinematical region where the decay channel amplitudes are valid are: $x = 1-y-z, \,\, 0 \leq y \leq 1-z, \,\, 0 \leq z$.
The amplitudes $|\mathcal{A}_f\rangle$ can be organized in terms of the operators and the colored final states. For $f=ggg$, there are two amplitudes, one for $O_{G}(x)$ and another for $O_{J}(x)$. When $f=q\bar{q}g$, one can construct two more amplitudes. However, not all of them start at ${\mathcal O}(g_s)$. Our interest is in the quantity
${\cal S}_{f} = \langle\mathcal{A}_f|\mathcal{A}
_f\rangle $; in order to do so, we perform the Dirac and SU(N) color manipulations with our in-house codes in FORM \cite{Ruijl:2017dtg}.
As a result we can express ${\cal S}_{f}$ in terms of rational coefficients and master intergrals, where the later has been taken from \cite{Gehrmann:2023etk}. They have been 
 analytically computed up to ${\mathcal O}(\epsilon^2)$, in terms of Generalised Polylogarithms (GPLs). $\epsilon$
 is the dimensional regularization parameter, used to regulate the ultraviolet  (UV) and infrared (IR) divergences.
During the intermediate stages of the calculations, the size of the polynomials becomes enormous, which results in final file sizes of $\sim$ a few GB.  Using MultivariateApart \cite{Heller:2021qkz}, we have simplified the Integration by Parts identities that appear in our computation.

\section{UV renormalization and IR regularization}
We regularize the ultraviolet(UV) and infrared singularities(IR) that arise in the matrix elements, using dimensional regularization ($d= 4 + \epsilon$). First we shall discuss about the UV renormalization.
Apart from the renormalization of the strong coupling constant, renormalization of the composite operators $O_{G}(x)$ and $O_{J}(x)$ needs also to be performed.
The choice of representation of $\gamma_5$ in Eq. \ref{eq:g5}  fails to satisfy the Ward identities.  In addition $\{ \gamma_{\mu},\gamma_{5}\} \neq 0$ in $d$ dimensions. One way to resolve the problem is to define the axial singlet current correctly, which is as follows \cite{Larin:1993tq}:
 \begin{equation}
 J_{\mu}^5 = \bar{\psi}  \gamma_{\mu} \gamma_{5}\psi
 \end{equation}
 The renormalization of the axial singlet current is first performed using a finite renormalization constant, $Z_5^{s}$ \cite{Trueman:1979en,Larin:1993tq}, followed by multiplying the overall operator renormalization constant, $Z^{s}_{\overline{MS}} $. The analytical expression of $Z^{s}_{\overline{MS}} $ till third order in the strong coupling constant can be found in \cite{Larin:1993tq,Zoller:2013ixa}, which was later verified independently in
 \cite{Ahmed:2015qpa}. The final expressions for the renormalized operators in terms of the bare counterparts are as follows:
 \begin{align}
  \label{eq:OJRen}
  \left[ O_{J} \right]_{R} &= Z^{s}_{5} Z^{s}_{\overline{MS}} \left[O_{J}\right]_{B}\equiv Z_{JJ}\left[O_{J}\right]_{B}\,, \nonumber \\
\left[ O_{G} \right]_{R} &= Z_{GG} \left[ O_{G}\right]_B +
  Z_{GJ} \left[ O_{J} \right]_B
\end{align}
After UV renormalization, the matrix amplitudes still contain divergences, arising from the IR sector, due to presence of massless particles. It is now well known that for a $n$-loop amplitude, the IR singularities factorize in the color space in terms of the lower $n-1$ loop amplitudes. Using the factorization and resummation properties of QCD amplitudes, the IR poles for a n-point 2 loop amplitude can be predicted, thanks to the seminal works in \cite{Catani:1998bh, Sterman:2002qn, Becher:2009cu}. The IR singular contributions from $S_{ggg}$ and $S_{q\bar{q}g}$ are same as the universal IR predictions in \cite{Catani:1998bh}. 
A comparison of the finite parts of $S_{ggg}$ and $S_{q\bar{q}g}$, between this study, to the one in ~\cite{Banerjee:2017faz}, for a few phase-space points is presented in Table~\ref{tab:matrix_check}. For the detailed expression of $S_{ggg}$ and $S_{q\bar{q}g}$ in terms of $C_G^{(1)}$ and $C_J^{(1)}$, see Eq. 19 of \cite{Banerjee:2024zho}.
The finite part of $S_{ggg}$ and $S_{q\bar{q}g}$ in   ~\cite{Banerjee:2017faz} has been 
confirmed in the recent article ~\cite{Kim:2024kaq}. 

\section{Results and discussions}
In \cite{Banerjee:2024zho}, we have provided a numerical implementation for the  two-loop amplitudes. For a fast numerical grid, we simplified the ${\mathcal O}(\epsilon^0)$-${\mathcal O}(\epsilon^2)$ results at two-loop and ${\mathcal O}(\epsilon^0)$-${\mathcal O}(\epsilon^4)$ results at one-loop, with our in-house routines in Mathematica.
We provide our numerical implementations in FORTRAN-95 routines, which can be used with any Monte Carlo phase space generator. The relavent numerical codes can be obtain from source files of Ref.~\cite{Banerjee:2024zho}. 
In any Monte-Carlo study, these results need to be integrated over the allowed phase space regions, as desired by experimental searches. The number of phase space points 
needed depends on the accuracy demanded for phenomenological studies. Thus it is essential to 
efficiently implement the huge analytical results, in a way that the time spent during the numerical runs in the computer algebra part is minimal. In \cite{Banerjee:2024zho} we have shown the effect of using different optimization routines in FORM on the evaluation time of the simplified matrix elements. It is to be noted that these optimizations depend on the type of matrix elements($q{\bar q}g $ or $ggg$) and also on the order of the perturbation theory. After the optimizations, compiling all the two-loop Fortran files takes $\sim$ 10 min on a 3.2 GHz computer, with 32 CPUs and 64 GB RAM.
These simplified matrix elements were then evaluated in different phase space regions, see Table I of \cite{Banerjee:2024zho}. Therein we observed that at NLO, the ${\mathcal O}(\epsilon^0)$ and ${\mathcal O}(\epsilon^1)$, takes $\sim$ 100 $\mu$s to run, per phase point. At  ${\mathcal O}(\epsilon^4)$, the running time for the matrix elements is about 1 second, for the sample phase-space points considered. For the phase space regions as we considered in our study, the NNLO matrix elements took $\sim$ 5 $\mu$s to run per phase space point at ${\mathcal O}(\epsilon^0)$, which went up to $\sim$ 19 s  at ${\mathcal O}(\epsilon^2)$. We also observed that the  ${\mathcal O}(\epsilon^2)$ contribution for ${ggg}$ channel is about 100 times large compared to ${q\bar{q}g}$ channel.

\begin{table}[h!]
\centering
\scriptsize
\setlength{\tabcolsep}{5pt}
\renewcommand{\arraystretch}{1.2}

\resizebox{\textwidth}{!}{
\begin{tabular}{c c c c c c c}
\hline
\textbf{Matrix elements} & \textbf{Contribution} & \textbf{Phase Space points} &
$\mathcal{O}(\epsilon^0)$  in \cite{Banerjee:2017faz} &
Our $\mathcal{O}(\epsilon^0)$ & Our $\mathcal{O}(\epsilon)$ & Our $\mathcal{O}(\epsilon^2)$ \\
\hline
\hline

%

$S_{ggg}$ & $\left(C_G^{(1)}\right)^2$ & $y=0.1, z=0.4$ &
$-2.577\times10^9 $&
$-2.573\times10^9 $& $3.426\times10^9$ & $3.979\times10^9$ \\

$S_{ggg}$ & $\left(C_G^{(1)}\right)^2$ & $y=0.3, z=0.5$ &
$-1.670\times10^9$ &
$-1.667\times10^9$ & $2.209\times10^9$ & $2.042\times10^9$ \\

%

$S_{q\bar{q}g}$  & $\left(C_G^{(1)}\right)^2$ & $y=0.1, z=0.4$ &
$-813958.263$ &
$-813958.263$ & $1.178\times10^7$ & $1.104\times10^7$ \\

$S_{q\bar{q}g}$ & $\left(C_G^{(1)}\right)^2$ & $y=0.3, z=0.5$ &
$-1.244\times10^7$ &
$-1.244\times10^7$ & $4.252\times10^7$  & $-2.239\times10^7$  \\
$S_{q\bar{q}g}$ & $\left(C_G^{(1)}\right)^2 C_J^{(1)}$ & $y=0.1, z=0.4$ &
$8704.0$ &
$8704.0$ & $-367529.788$ & $-12807.054$ \\

$S_{q\bar{q}g}$ & $\left(C_G^{(1)}\right)^2 C_J^{(1)}$ & $y=0.3, z=0.5$ &
$43520.0$ &
$43520.0$ & $-118615.169$ & $-19538.771$ \\
\hline
\end{tabular}
}

\caption{Comparison(in the 4th and 5th column) of the matrix elements, at two loops, between this calculation and that of Ref.~\cite{Banerjee:2017faz} for two different phase-space points at $Q^2=\mu_R^2=10$. The 6th and 7th column contains 
higher powers of matrix elements beyond two loops, for the same phase space points.}
\label{tab:matrix_check}

\end{table}

\section{Summary}
In this talk, we have presented, for the first time, analytically computed matrix elements for the decay of a pseudo-scalar Higgs boson into $ggg$ and $q\bar{q}g$, which contribute beyond NNLO in QCD. These results will be important for precise predictions for pseudo-scalar Higgs production in association with a jet beyond NNLO.

%

%

\begin{thebibliography}{99}
\makeatletter
\providecommand \@ifxundefined [1]{%
 \@ifx{#1\undefined}
}%
\providecommand \@ifnum [1]{%
 \ifnum #1\expandafter \@firstoftwo
 \else \expandafter \@secondoftwo
 \fi
}%
\providecommand \@ifx [1]{%
 \ifx #1\expandafter \@firstoftwo
 \else \expandafter \@secondoftwo
 \fi
}%
\providecommand \natexlab [1]{#1}%
\providecommand \enquote  [1]{``#1''}%
\providecommand \bibnamefont  [1]{#1}%
\providecommand \bibfnamefont [1]{#1}%
\providecommand \citenamefont [1]{#1}%
\providecommand \href@noop [0]{\@secondoftwo}%
\providecommand \href [0]{\begingroup \@sanitize@url \@href}%
\providecommand \@href[1]{\@@startlink{#1}\@@href}%
\providecommand \@@href[1]{\endgroup#1\@@endlink}%
\providecommand \@sanitize@url [0]{\catcode `\\12\catcode `\$12\catcode `\&12\catcode `\#12\catcode `\^12\catcode `\_12\catcode `\%12\relax}%
\providecommand \@@startlink[1]{}%
\providecommand \@@endlink[0]{}%
\providecommand \url  [0]{\begingroup\@sanitize@url \@url }%
\providecommand \@url [1]{\endgroup\@href {#1}{\urlprefix }}%
\providecommand \urlprefix  [0]{URL }%
\providecommand \Eprint [0]{\href }%
\providecommand \doibase [0]{http://dx.doi.org/}%
\providecommand \selectlanguage [0]{\@gobble}%
\providecommand \bibinfo  [0]{\@secondoftwo}%
\providecommand \bibfield  [0]{\@secondoftwo}%
\providecommand \translation [1]{[#1]}%
\providecommand \BibitemOpen [0]{}%
\providecommand \bibitemStop [0]{}%
\providecommand \bibitemNoStop [0]{.\EOS\space}%
\providecommand \EOS [0]{\spacefactor3000\relax}%
\providecommand \BibitemShut  [1]{\csname bibitem#1\endcsname}%
\let\auto@bib@innerbib\@empty
\bibitem [{\citenamefont {Aad}\ \emph {et~al.}(2012)\citenamefont {Aad} \emph {et~al.}}]{ATLAS:2012yve}%
  \BibitemOpen
  \bibfield  {author} {\bibinfo {author} {\bibfnamefont {G.}~\bibnamefont {Aad}} \emph {et~al.} (\bibinfo {collaboration} {ATLAS}),\ }\href {\doibase 10.1016/j.physletb.2012.08.020} {\bibfield  {journal} {\bibinfo  {journal} {Phys. Lett. B}\ }\textbf {\bibinfo {volume} {716}},\ \bibinfo {pages} {1} (\bibinfo {year} {2012})},\ \Eprint {http://arxiv.org/abs/1207.7214} {arXiv:1207.7214 [hep-ex]} \BibitemShut {NoStop}%
\bibitem [{\citenamefont {Chatrchyan}\ \emph {et~al.}(2012)\citenamefont {Chatrchyan} \emph {et~al.}}]{CMS:2012qbp}%
  \BibitemOpen
  \bibfield  {author} {\bibinfo {author} {\bibfnamefont {S.}~\bibnamefont {Chatrchyan}} \emph {et~al.} (\bibinfo {collaboration} {CMS}),\ }\href {\doibase 10.1016/j.physletb.2012.08.021} {\bibfield  {journal} {\bibinfo  {journal} {Phys. Lett. B}\ }\textbf {\bibinfo {volume} {716}},\ \bibinfo {pages} {30} (\bibinfo {year} {2012})},\ \Eprint {http://arxiv.org/abs/1207.7235} {arXiv:1207.7235 [hep-ex]} \BibitemShut {NoStop}%
  \bibitem [{\citenamefont {Ravindran}\ \emph {et~al.}(2002)\citenamefont {Ravindran}, \citenamefont {Smith},\ and\ \citenamefont {Van~Neerven}}]{Ravindran:2002dc}%
  \BibitemOpen
  \bibfield  {author} {\bibinfo {author} {\bibfnamefont {V.}~\bibnamefont {Ravindran}}, \bibinfo {author} {\bibfnamefont {J.}~\bibnamefont {Smith}}, \ and\ \bibinfo {author} {\bibfnamefont {W.~L.}\ \bibnamefont {Van~Neerven}},\ }\href {\doibase 10.1016/S0550-3213(02)00333-4} {\bibfield  {journal} {\bibinfo  {journal} {Nucl. Phys. B}\ }\textbf {\bibinfo {volume} {634}},\ \bibinfo {pages} {247} (\bibinfo {year} {2002})},\ \Eprint {http://arxiv.org/abs/hep-ph/0201114} {arXiv:hep-ph/0201114} \BibitemShut {NoStop}%
\bibitem [{\citenamefont {Harlander}\ and\ \citenamefont {Kilgore}(2002)}]{Harlander:2002vv}%
 \BibitemOpen
 \bibfield  {author} {\bibinfo {author} {\bibfnamefont {R.~V.}\ \bibnamefont {Harlander}}\ and\ \bibinfo {author} {\bibfnamefont {W.~B.}\ \bibnamefont {Kilgore}},\ }\href {\doibase 10.1088/1126-6708/2002/10/017} {\bibfield  {journal} {\bibinfo  {journal} {JHEP}\ }\textbf {\bibinfo {volume} {10}},\ \bibinfo {pages} {017} (\bibinfo {year} {2002})},\ \Eprint {http://arxiv.org/abs/hep-ph/0208096} {arXiv:hep-ph/0208096 [hep-ph]} \BibitemShut {NoStop}%
\bibitem [{\citenamefont {Ravindran}\ \emph {et~al.}(2003)\citenamefont {Ravindran}, \citenamefont {Smith},\ and\ \citenamefont {van Neerven}}]{Ravindran:2003um}%
 \BibitemOpen
 \bibfield  {author} {\bibinfo {author} {\bibfnamefont {V.}~\bibnamefont {Ravindran}}, \bibinfo {author} {\bibfnamefont {J.}~\bibnamefont {Smith}}, \ and\ \bibinfo {author} {\bibfnamefont {W.~L.}\ \bibnamefont {van Neerven}},\ }\href {\doibase 10.1016/S0550-3213(03)00457-7} {\bibfield  {journal} {\bibinfo  {journal} {Nucl. Phys.}\ }\textbf {\bibinfo {volume} {B665}},\ \bibinfo {pages} {325} (\bibinfo {year} {2003})},\ \Eprint {http://arxiv.org/abs/hep-ph/0302135} {arXiv:hep-ph/0302135 [hep-ph]} \BibitemShut {NoStop}%
 \bibitem [{\citenamefont {Anastasiou}\ \emph {et~al.}(2015)\citenamefont {Anastasiou}, \citenamefont {Duhr}, \citenamefont {Dulat}, \citenamefont {Herzog},\ and\ \citenamefont {Mistlberger}}]{Anastasiou:2015vya}%
 \BibitemOpen
 \bibfield  {author} {\bibinfo {author} {\bibfnamefont {C.}~\bibnamefont {Anastasiou}}, \bibinfo {author} {\bibfnamefont {C.}~\bibnamefont {Duhr}}, \bibinfo {author} {\bibfnamefont {F.}~\bibnamefont {Dulat}}, \bibinfo {author} {\bibfnamefont {F.}~\bibnamefont {Herzog}}, \ and\ \bibinfo {author} {\bibfnamefont {B.}~\bibnamefont {Mistlberger}},\ }\href {\doibase 10.1103/PhysRevLett.114.212001} {\bibfield  {journal} {\bibinfo  {journal} {Phys. Rev. Lett.}\ }\textbf {\bibinfo {volume} {114}},\ \bibinfo {pages} {212001} (\bibinfo {year} {2015})},\ \Eprint {http://arxiv.org/abs/1503.06056} {arXiv:1503.06056 [hep-ph]} \BibitemShut {NoStop}%
\bibitem [{\citenamefont {Anastasiou}\ \emph {et~al.}(2016)\citenamefont {Anastasiou}, \citenamefont {Duhr}, \citenamefont {Dulat}, \citenamefont {Furlan}, \citenamefont {Gehrmann}, \citenamefont {Herzog}, \citenamefont {Lazopoulos},\ and\ \citenamefont {Mistlberger}}]{Anastasiou:2016cez}%
 \BibitemOpen
 \bibfield  {author} {\bibinfo {author} {\bibfnamefont {C.}~\bibnamefont {Anastasiou}}, \bibinfo {author} {\bibfnamefont {C.}~\bibnamefont {Duhr}}, \bibinfo {author} {\bibfnamefont {F.}~\bibnamefont {Dulat}}, \bibinfo {author} {\bibfnamefont {E.}~\bibnamefont {Furlan}}, \bibinfo {author} {\bibfnamefont {T.}~\bibnamefont {Gehrmann}}, \bibinfo {author} {\bibfnamefont {F.}~\bibnamefont {Herzog}}, \bibinfo {author} {\bibfnamefont {A.}~\bibnamefont {Lazopoulos}}, \ and\ \bibinfo {author} {\bibfnamefont {B.}~\bibnamefont {Mistlberger}},\ }\href {\doibase 10.1007/JHEP05(2016)058} {\bibfield  {journal} {\bibinfo  {journal} {JHEP}\ }\textbf {\bibinfo {volume} {05}},\ \bibinfo {pages} {058} (\bibinfo {year} {2016})},\ \Eprint {http://arxiv.org/abs/1602.00695} {arXiv:1602.00695 [hep-ph]} \BibitemShut {NoStop}%
 \bibitem [{\citenamefont {Fayet}(1975)}]{Fayet:1974pd}%
  \BibitemOpen
  \bibfield  {author} {\bibinfo {author} {\bibfnamefont {P.}~\bibnamefont {Fayet}},\ }\href {\doibase 10.1016/0550-3213(75)90636-7} {\bibfield  {journal} {\bibinfo  {journal} {Nucl. Phys. B}\ }\textbf {\bibinfo {volume} {90}},\ \bibinfo {pages} {104} (\bibinfo {year} {1975})}\BibitemShut {NoStop}%
\bibitem [{\citenamefont {de~Florian}\ \emph {et~al.}(1999)\citenamefont {de~Florian}, \citenamefont {Grazzini},\ and\ \citenamefont {Kunszt}}]{deFlorian:1999zd}%
  \BibitemOpen
  \bibfield  {author} {\bibinfo {author} {\bibfnamefont {D.}~\bibnamefont {de~Florian}}, \bibinfo {author} {\bibfnamefont {M.}~\bibnamefont {Grazzini}}, \ and\ \bibinfo {author} {\bibfnamefont {Z.}~\bibnamefont {Kunszt}},\ }\href {\doibase 10.1103/PhysRevLett.82.5209} {\bibfield  {journal} {\bibinfo  {journal} {Phys. Rev. Lett.}\ }\textbf {\bibinfo {volume} {82}},\ \bibinfo {pages} {5209} (\bibinfo {year} {1999})},\ \Eprint {http://arxiv.org/abs/hep-ph/9902483} {arXiv:hep-ph/9902483} \BibitemShut {NoStop}%
  \bibitem [{\citenamefont {Field}\ \emph {et~al.}(2003)\citenamefont {Field}, \citenamefont {Smith}, \citenamefont {Tejeda-Yeomans},\ and\ \citenamefont {van Neerven}}]{Field:2002pb}%
  \BibitemOpen
  \bibfield  {author} {\bibinfo {author} {\bibfnamefont {B.}~\bibnamefont {Field}}, \bibinfo {author} {\bibfnamefont {J.}~\bibnamefont {Smith}}, \bibinfo {author} {\bibfnamefont {M.~E.}\ \bibnamefont {Tejeda-Yeomans}}, \ and\ \bibinfo {author} {\bibfnamefont {W.~L.}\ \bibnamefont {van Neerven}},\ }\href {\doibase 10.1016/S0370-2693(02)03048-4} {\bibfield  {journal} {\bibinfo  {journal} {Phys. Lett. B}\ }\textbf {\bibinfo {volume} {551}},\ \bibinfo {pages} {137} (\bibinfo {year} {2003})},\ \Eprint {http://arxiv.org/abs/hep-ph/0210369} {arXiv:hep-ph/0210369} \BibitemShut {NoStop}%
\bibitem [{\citenamefont {Bernreuther}\ \emph {et~al.}(2010)\citenamefont {Bernreuther}, \citenamefont {Gonzalez},\ and\ \citenamefont {Wiebusch}}]{Bernreuther:2010uw}%
  \BibitemOpen
  \bibfield  {author} {\bibinfo {author} {\bibfnamefont {W.}~\bibnamefont {Bernreuther}}, \bibinfo {author} {\bibfnamefont {P.}~\bibnamefont {Gonzalez}}, \ and\ \bibinfo {author} {\bibfnamefont {M.}~\bibnamefont {Wiebusch}},\ }\href {\doibase 10.1140/epjc/s10052-010-1335-1} {\bibfield  {journal} {\bibinfo  {journal} {Eur. Phys. J. C}\ }\textbf {\bibinfo {volume} {69}},\ \bibinfo {pages} {31} (\bibinfo {year} {2010})},\ \Eprint {http://arxiv.org/abs/1003.5585} {arXiv:1003.5585 [hep-ph]} \BibitemShut {NoStop}%
  \bibitem [{\citenamefont {Kim}\ and\ \citenamefont {Williams}(2024)}]{Kim:2024kaq}%
  \BibitemOpen
  \bibfield  {author} {\bibinfo {author} {\bibfnamefont {Y.}~\bibnamefont {Kim}}\ and\ \bibinfo {author} {\bibfnamefont {C.}~\bibnamefont {Williams}},\ }\href {\doibase 10.1007/JHEP08(2024)042} {\bibfield  {journal} {\bibinfo  {journal} {JHEP}\ }\textbf {\bibinfo {volume} {08}},\ \bibinfo {pages} {042} (\bibinfo {year} {2024})},\ \BibitemShut {NoStop}%
  \bibitem [{\citenamefont {Gehrmann}\ \emph {et~al.}(2023)\citenamefont {Gehrmann}, \citenamefont {Jakub\v{c}\'\i{}k}, \citenamefont {Mella}, \citenamefont {Syrrakos},\ and\ \citenamefont {Tancredi}}]{Gehrmann:2023etk}%
  \BibitemOpen
  \bibfield  {author} {\bibinfo {author} {\bibfnamefont {T.}~\bibnamefont {Gehrmann}}, \bibinfo {author} {\bibfnamefont {P.}~\bibnamefont {Jakub\v{c}\'\i{}k}}, \bibinfo {author} {\bibfnamefont {C.~C.}\ \bibnamefont {Mella}}, \bibinfo {author} {\bibfnamefont {N.}~\bibnamefont {Syrrakos}}, \ and\ \bibinfo {author} {\bibfnamefont {L.}~\bibnamefont {Tancredi}},\ }\href {\doibase 10.1007/JHEP04(2023)016} {\bibfield  {journal} {\bibinfo  {journal} {JHEP}\ }\textbf {\bibinfo {volume} {04}},\ \bibinfo {pages} {016} (\bibinfo {year} {2023})}\BibitemShut {NoStop}%
  \bibitem [{\citenamefont {Chetyrkin}\ \emph {et~al.}(1998)\citenamefont {Chetyrkin}, \citenamefont {Kniehl}, \citenamefont {Steinhauser},\ and\ \citenamefont {Bardeen}}]{Chetyrkin:1998mw}%
  \BibitemOpen
  \bibfield  {author} {\bibinfo {author} {\bibfnamefont {K.~G.}\ \bibnamefont {Chetyrkin}}, \bibinfo {author} {\bibfnamefont {B.~A.}\ \bibnamefont {Kniehl}}, \bibinfo {author} {\bibfnamefont {M.}~\bibnamefont {Steinhauser}}, \ and\ \bibinfo {author} {\bibfnamefont {W.~A.}\ \bibnamefont {Bardeen}},\ }\href {\doibase 10.1016/S0550-3213(98)00594-X} {\bibfield  {journal} {\bibinfo  {journal} {Nucl. Phys.}\ }\textbf {\bibinfo {volume} {B535}},\ \bibinfo {pages} {3} (\bibinfo {year} {1998})},\ \Eprint {http://arxiv.org/abs/hep-ph/9807241} {arXiv:hep-ph/9807241 [hep-ph]} \BibitemShut {NoStop}%
  \bibitem [{\citenamefont {Banerjee}\ \emph {et~al.}(2025)\citenamefont {Banerjee}, \citenamefont {Dey}, \citenamefont {Kumar},\ and\ \citenamefont {Ravindran}}]{Banerjee:2024zho}%
  \BibitemOpen
  \bibfield  {author} {\bibinfo {author} {\bibfnamefont {P.}~\bibnamefont {Banerjee}}, \bibinfo {author} {\bibfnamefont {C.}~\bibnamefont {Dey}}, \bibinfo {author} {\bibfnamefont {M.~C.}\ \bibnamefont {Kumar}}, \ and\ \bibinfo {author} {\bibfnamefont {V.}~\bibnamefont {Ravindran}},\ }\href {\doibase 10.1103/PhysRevD.111.054037} {\bibfield  {journal} {\bibinfo  {journal} {Phys. Rev. D}\ }\textbf {\bibinfo {volume} {111}},\ \bibinfo {pages} {054037} (\bibinfo {year} {2025})} \BibitemShut {NoStop}%
\bibitem [{\citenamefont {'t~Hooft}\ and\ \citenamefont {Veltman}(1972)}]{tHooft:1972tcz}%
  \BibitemOpen
  \bibfield  {author} {\bibinfo {author} {\bibfnamefont {G.}~\bibnamefont {'t~Hooft}}\ and\ \bibinfo {author} {\bibfnamefont {M.~J.~G.}\ \bibnamefont {Veltman}},\ }\href {\doibase 10.1016/0550-3213(72)90279-9} {\bibfield  {journal} {\bibinfo  {journal} {Nucl. Phys.}\ }\textbf {\bibinfo {volume} {B44}},\ \bibinfo {pages} {189} (\bibinfo {year} {1972})}\BibitemShut {NoStop}
\bibitem [{\citenamefont {Ruijl}\ \emph {et~al.}(2017)\citenamefont {Ruijl}, \citenamefont {Ueda},\ and\ \citenamefont {Vermaseren}}]{Ruijl:2017dtg}%
  \BibitemOpen
  \bibfield  {author} {\bibinfo {author} {\bibfnamefont {B.}~\bibnamefont {Ruijl}}, \bibinfo {author} {\bibfnamefont {T.}~\bibnamefont {Ueda}}, \ and\ \bibinfo {author} {\bibfnamefont {J.}~\bibnamefont {Vermaseren}},\ }\href@noop {} {\  (\bibinfo {year} {2017})}\BibitemShut {NoStop}%
\bibitem [{\citenamefont {Heller}\ and\ \citenamefont {von Manteuffel}(2022)}]{Heller:2021qkz}%
  \BibitemOpen
  \bibfield  {author} {\bibinfo {author} {\bibfnamefont {M.}~\bibnamefont {Heller}}\ and\ \bibinfo {author} {\bibfnamefont {A.}~\bibnamefont {von Manteuffel}},\ }\href {\doibase 10.1016/j.cpc.2021.108174} {\bibfield  {journal} {\bibinfo  {journal} {Comput. Phys. Commun.}\ }\textbf {\bibinfo {volume} {271}},\ \bibinfo {pages} {108174} (\bibinfo {year} {2022})}\BibitemShut {NoStop}%
%
\bibitem [{\citenamefont {Larin}(1993)}]{Larin:1993tq}%
  \BibitemOpen
  \bibfield  {author} {\bibinfo {author} {\bibfnamefont {S.~A.}\ \bibnamefont {Larin}},\ }\href {\doibase 10.1016/0370-2693(93)90053-K} {\bibfield  {journal} {\bibinfo  {journal} {Phys. Lett.}\ }\textbf {\bibinfo {volume} {B303}},\ \bibinfo {pages} {113} (\bibinfo {year} {1993})},\ \BibitemShut {NoStop}%
%
\bibitem [{\citenamefont {Trueman}(1979)}]{Trueman:1979en}%
  \BibitemOpen
  \bibfield  {author} {\bibinfo {author} {\bibfnamefont {T.~L.}\ \bibnamefont {Trueman}},\ }\href {\doibase 10.1016/0370-2693(79)90480-5} {\bibfield  {journal} {\bibinfo  {journal} {Phys. Lett. B}\ }\textbf {\bibinfo {volume} {88}},\ \bibinfo {pages} {331} (\bibinfo {year} {1979})}\BibitemShut {NoStop}%
%
\bibitem [{\citenamefont {Zoller}(2013)}]{Zoller:2013ixa}%
  \BibitemOpen
  \bibfield  {author} {\bibinfo {author} {\bibfnamefont {M.~F.}\ \bibnamefont {Zoller}},\ }\href {\doibase 10.1007/JHEP07(2013)040} {\bibfield  {journal} {\bibinfo  {journal} {JHEP}\ }\textbf {\bibinfo {volume} {07}},\ \bibinfo {pages} {040} (\bibinfo {year} {2013})},\  \BibitemShut {NoStop}%
%
\bibitem [{\citenamefont {Ahmed}\ \emph {et~al.}(2015)\citenamefont {Ahmed}, \citenamefont {Gehrmann}, \citenamefont {Mathews}, \citenamefont {Rana},\ and\ \citenamefont {Ravindran}}]{Ahmed:2015qpa}%
  \BibitemOpen
  \bibfield  {author} {\bibinfo {author} {\bibfnamefont {T.}~\bibnamefont {Ahmed}}, \bibinfo {author} {\bibfnamefont {T.}~\bibnamefont {Gehrmann}}, \bibinfo {author} {\bibfnamefont {P.}~\bibnamefont {Mathews}}, \bibinfo {author} {\bibfnamefont {N.}~\bibnamefont {Rana}}, \ and\ \bibinfo {author} {\bibfnamefont {V.}~\bibnamefont {Ravindran}},\ }\href {\doibase 10.1007/JHEP11(2015)169} {\bibfield  {journal} {\bibinfo  {journal} {JHEP}\ }\textbf {\bibinfo {volume} {11}},\ \bibinfo {pages} {169} (\bibinfo {year} {2015})},\  \BibitemShut {NoStop}%
\bibitem [{\citenamefont {Catani}(1998)}]{Catani:1998bh}%
  \BibitemOpen
  \bibfield  {author} {\bibinfo {author} {\bibfnamefont {S.}~\bibnamefont {Catani}},\ }\href {\doibase 10.1016/S0370-2693(98)00332-3} {\bibfield  {journal} {\bibinfo  {journal} {Phys. Lett.}\ }\textbf {\bibinfo {volume} {B427}},\ \bibinfo {pages} {161} (\bibinfo {year} {1998})},\ \BibitemShut {NoStop}%
  \bibitem [{\citenamefont {Sterman}\ and\ \citenamefont {Tejeda-Yeomans}(2003)}]{Sterman:2002qn}%
  \BibitemOpen
  \bibfield  {author} {\bibinfo {author} {\bibfnamefont {G.~F.}\ \bibnamefont {Sterman}}\ and\ \bibinfo {author} {\bibfnamefont {M.~E.}\ \bibnamefont {Tejeda-Yeomans}},\ }\href {\doibase 10.1016/S0370-2693(02)03100-3} {\bibfield  {journal} {\bibinfo  {journal} {Phys. Lett.}\ }\textbf {\bibinfo {volume} {B552}},\ \bibinfo {pages} {48} (\bibinfo {year} {2003})},\ \BibitemShut {NoStop}%
\bibitem [{\citenamefont {Becher}\ and\ \citenamefont {Neubert}(2009)}]{Becher:2009cu}%
  \BibitemOpen
  \bibfield  {author} {\bibinfo {author} {\bibfnamefont {T.}~\bibnamefont {Becher}}\ and\ \bibinfo {author} {\bibfnamefont {M.}~\bibnamefont {Neubert}},\ }\href {\doibase 10.1103/PhysRevLett.102.162001, 10.1103/PhysRevLett.111.199905} {\bibfield  {journal} {\bibinfo  {journal} {Phys. Rev. Lett.}\ }\textbf {\bibinfo {volume} {102}},\ \bibinfo {pages} {162001} (\bibinfo {year} {2009})},\ \bibinfo {note} {[Erratum: Phys. Rev. Lett.111,no.19,199905(2013)]},\ \BibitemShut {NoStop}%
\bibitem [{\citenamefont {Banerjee}\ \emph {et~al.}(2017)\citenamefont {Banerjee}, \citenamefont {Dhani},\ and\ \citenamefont {Ravindran}}]{Banerjee:2017faz}%
  \BibitemOpen
  \bibfield  {author} {\bibinfo {author} {\bibfnamefont {P.}~\bibnamefont {Banerjee}}, \bibinfo {author} {\bibfnamefont {P.~K.}\ \bibnamefont {Dhani}}, \ and\ \bibinfo {author} {\bibfnamefont {V.}~\bibnamefont {Ravindran}},\ }\href {\doibase 10.1007/JHEP10(2017)067} {\bibfield  {journal} {\bibinfo  {journal} {JHEP}\ }\textbf {\bibinfo {volume} {10}},\ \bibinfo {pages} {067} (\bibinfo {year} {2017})},\ \BibitemShut {NoStop}%
%
\end{thebibliography}
\end{document}